\documentclass[twocolumn,showpacs,preprintnumbers,amsmath,amssymb]{revtex4}
\usepackage{graphicx}
\usepackage{dcolumn}
\usepackage{bm}
\usepackage{epsfig}
\newcommand{\QKD}{{\sc QKD}}

\newcommand{\WCP}{{\sc WCP}}

\newcommand{\ket}[1]{\mbox{$ | #1 \rangle $}}
\newcommand{\bra}[1]{\mbox{$ \langle #1 | $}}

\newcommand{\one}{\leavevmode\hbox{\small1\normalsize\kern-.33em1}}

\begin{document}

\title{Intercept-resend attacks in the Bennett-Brassard 1984 quantum key distribution protocol with weak coherent pulses}
\author{Marcos Curty and Norbert L\"{u}tkenhaus}
\affiliation{Quantum Information Theory Group, Institut f\"ur
Theoretische Physik I, and Max-Planck Research Group, Institute
of Optics, Information and Photonics, Universit\"{a}t
Erlangen-N\"{u}rnberg, Staudtstra{\ss}e 7/B2, 91058 Erlangen,
Germany}

\begin{abstract}
Unconditional security proofs of the Bennett-Brassard protocol of
quantum key distribution have been obtained recently. These
proofs cover also practical implementations that utilize weak
coherent pulses in the four signal polarizations. Proven secure
rates leave open the possibility that new proofs or new public
discussion protocols obtain larger rates over increased distance.
In this paper we investigate limits to error rate and signal
losses that can be tolerated by future protocols and proofs.
\end{abstract}
\maketitle

\section{INTRODUCTION}
Quantum key distribution (\QKD) \cite{Wiesner83,BB84} is a
technique that exploits quantum effects to establish a secure
secret key between two parties (usually called Alice and Bob).
This secret key is the essential ingredient of the one-time-pad
or Vernam cipher \cite{Vernam26}, the only known encryption
method that can provide information-theoretic secure
communications.

The first complete \QKD\ scheme is that introduced by Bennett and
Brassard in $1984$ (BB$84$ for short) \cite{BB84}. In a quantum
optical implementation of this protocol, Alice encodes each
random bit into the polarization state of a single-photon: she
chooses along one of two mutually unbiased bases, e.g. either a
linear or a circular polarization basis. On the receiving side,
Bob measures each photon by selecting at random between two
polarization analyzers, one for each possible basis. As a result,
Alice and Bob end up with some classical correlated data that can
be described by a joint probability distribution $P(A,B)$, where
the random variables $A$ and $B$ represent the signal states
prepared by Alice and the measurement results obtained by Bob,
respectively. Next, Alice and Bob use an authenticated public
channel to process these data in order to obtain a secret key.
This second phase, usually called {\it key distillation},
involves, typically, postselection of data, error correction to
reconcile the data, and privacy amplification to decouple the
data from a possible eavesdropper (Eve) \cite{Norbert99}. A full
proof of the unconditional security for the complete BB$84$
protocol has been obtained \cite{Mayers98}.

After the first experimental demonstration of the feasibility of
the BB$84$ scheme \cite{Bennett92}, several experimental groups
have realized long-distance implementations of QKD in the last
years \cite{expQKD}. However, these practical approaches differ
in many important aspects from the original theoretical proposal,
since that demands technologies that are beyond our present
experimental capability. Especially, the signal states emitted by
the source, instead of being single-photon signals, are usually
weak coherent pulses (\WCP) with a low probability of containing
more than one photon (typical average photon-numbers are $0.1$ or
higher). The quantum channel (e.g. optical fiber) introduces
considerable attenuation and errors that affect the signals even
when Eve is not present. Finally, the detectors employed by the
receiver have a low detection efficiency and are noisy due to
dark counts. All these modifications from the ideal BB$84$
protocol to real implementations can jeopardize the security of
the protocol, and lead to limitations of rate and distance that
can be covered by these techniques \cite{Huttner95,Nor00}. A
positive security proof against all individual particle attacks,
even with practical signals, has been given in \cite{Norbert00b}.
More recently, a complete proof of the unconditional security of
the BB$84$ scheme in a realistic setting has also been achieved
\cite{Inamori01}. This means that, despite of practical
imperfections, it is still possible to obtain a secure secret key
with the support of the classical information techniques used in
the key distillation phase.

While all these positive security proofs are of great importance
for QKD, they might be over-restrictive. This might be so either
because of the particular mathematical techniques used in the
security proofs, or because of the specific key distillation
protocols considered. Different ideas to extend the proven secure
regimes of Ref.~\cite{Mayers98,Inamori01} have been found
\cite{NoteGisin,NoteGisinb}. More recently, it has been shown that
the ultimate limit for secure QKD is given by the proven presence
of quantum correlations in $P(A,B)$ \cite{Curty04}: whenever the
joint probability distribution $P(A,B)$, together with the
knowledge of the corresponding signal states and detection
methods employed by Alice and Bob, can be interpreted as coming
from a separable state then no secret key can be obtained,
whatever classical protocol may be used in the key distillation
phase.

In this paper we are interested in this ultimate limit, given by
the existence of quantum correlations in $P(A,B)$, for practical
\QKD. In principle, to decide whether the joint probability
distribution $P(A,B)$, obtained after the first phase of QKD,
contains quantum correlations, Alice and Bob can use all the
information contained in $P(A,B)$. However, we evaluate only
those events where the signal preparation and detection methods
employed by Alice and Bob use the same polarization basis.
Moreover, in this paper, we consider that Alice and Bob analyze
only two particular events: The expected click rate at Bob's
side, and the generalized error rate. For this, we investigate a
simple eavesdropping strategy based on {\it intercept-resend}
attacks: Eve measures out every signal emitted by Alice and
prepares a new one, depending on the result obtained, that is
given to Bob. Intercept-resend attacks correspond to entanglement
breaking channels \cite{Horodecki03} and, therefore, they cannot
lead to a secure key \cite{Curty04}. This kind of attacks has
been already studied by Du$\check{s}$ek {\it et al.} in
Ref.~\cite{jahma01} and by F\'elix {\it et al.} in
Ref.~\cite{Felix01}. In particular, Du$\check{s}$ek {\it et al.}
considered the case where Eve realizes unambiguous state
discrimination (USD) \cite{usd,chef} of Alice's signal states.
Whenever Eve is successful with her USD measurement and she
identifies the signal state sent by Alice, then she sends this
information to a quantum source close to Bob via a classical
channel. This source prepares the identified quantum state and
subsequently forwards it to Bob. If the identification process
does not succeed, then Eve sends the vacuum signal to Bob to
avoid errors. This attack does not introduce errors in the signal
states, but it requires high losses in the channel. Note that in
order to unambiguously discriminate between the four BB$84$
signal states Eve requires that the signals contain, at least,
three or more photons \cite{chef,jahma01}. In this paper, we
investigate a different regime. Specifically, we propose a simple
intercept-resend attack for the scenario where the attenuation
introduced by the channel is not sufficiently high for Eve to
perform the USD attack. This fact is compensated by allowing her
to introduce some errors. As a result, we obtain an upper bound on
the maximal distance achievable by the BB$84$ protocol as a
function of the error rate and the mean photon-number of the
signals sent by Alice. Beyond this upper bound no secure QKD is
possible.

The paper is organized as follows. In Sec.~\ref{sec:SIGNALS} we
describe in more detail the signal states and detection methods
employed by Alice and Bob. Then, in Sec.~\ref{IR} we introduce an
intercept-resend attack for practical \QKD\ and we derive an
upper bound on the maximal distance achievable by the BB$84$
protocol. Finally, Sec.~\ref{CONC} concludes the paper with a
summary.

\section{TOOLBOX FOR ALICE AND BOB}\label{sec:SIGNALS}

\subsection{Alice}
We consider that Alice uses \WCP\ signal states described by
coherent states with a small amplitude $\alpha$. Moreover, we
assume the typical scenario where there is no phase reference
available outside Alice's setup. This results in effective signal
states which are mixtures of Fock states with a Poissonian
photon-number distribution of mean $\mu=|\alpha|^2$. Such states
are described as
\begin{equation}\label{SigAlice}
\rho_k=e^{-\mu}\ \sum_{n=0}^{\infty}\
\frac{\mu^n}{n!}\ket{n_{k}}\bra{n_{k}},
\end{equation}
where the states $\ket{n_{k}}$ denote Fock states with $n$ photons
in one of the four BB$84$ polarization states, which are labeled
with the index $k$, with $k=0, ..., 3$.

The symmetry imposed by the BB$84$ protocol on the signal states
$\ket{n_{k}}$, with $n>0$, guarantees that there exists a unitary
transformation $U(n)$ such that
\begin{eqnarray}
\ket{n_{k}}&=&U(n)\ket{n_{k-1}}=U^k(n)\ket{n_{0}}\nonumber\\
\ket{n_{0}}&=&U(n)\ket{n_{3}}\nonumber\\
U^4(n)&=&\openone.
\end{eqnarray}
For more details in our specific case see \cite{jahma01}.
The unitary operator $U(n)$ can be expanded as
\begin{equation}\label{uni_new}
U(n)=\sum_{j=0}^3 \exp
\bigg(2\pi{}i\frac{j}{4}\bigg)\ket{\phi_j(n)}\bra{\phi_j(n)},
\end{equation}
where the states $\ket{\phi_j(n)}$ represent a set of orthogonal
states. This allows us to write the states $\ket{n_{k}}$, with
$n>0$, as \cite{chef,ban}
\begin{equation}\label{dec}
\ket{n_k}=\sum_{j=0}^{3} c_j(n) \exp \bigg(2\pi{}i\frac{k
j}{4}\bigg)\ket{\phi_j(n)},
\end{equation}
where the coefficients $c_j(n)$ satisfy $\sum_j |c_j(n)|^2=1$.
The exact values of the coefficients $|c_j(n)|$ can be obtained
explicitly using the overlaps of the four states $\ket{n_k}$,
with $k=0, ..., 3$, according to the formula \cite{chef}
\begin{equation}
|c_j(n)|^2=\frac{1}{16}\sum_{l,m} \exp{\Bigg(-2\pi{}i\
\frac{j(l-m)}{4}\Bigg)}\bra{n_l}n_m\rangle.
\end{equation}
As a result, one finds the  expressions \cite{jahma01}
\begin{eqnarray}\label{coef}
|c_0(n)|&=&\sqrt{\frac{1}{4}+2^{-(1+n/2)}\cos\Big(\frac{\pi}{4}n\Big)},\\
|c_1(n)|&=&\sqrt{\frac{1}{4}+2^{-(1+n/2)}\sin\Big(\frac{\pi}{4}n\Big)},\nonumber\\
|c_2(n)|&=&\sqrt{\frac{1}{4}-2^{-(1+n/2)}\cos\Big(\frac{\pi}{4}n\Big)},\nonumber\\
|c_3(n)|&=&\sqrt{\frac{1}{4}-2^{-(1+n/2)}\sin\Big(\frac{\pi}{4}n\Big)}\nonumber.
\end{eqnarray}

\subsection{Bob}
\begin{figure}
\begin{center}
\includegraphics[scale=.9]{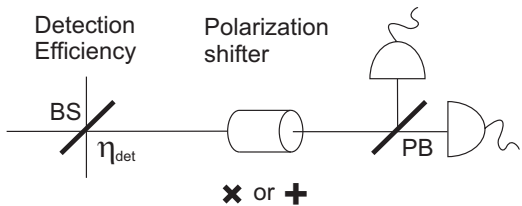}
\end{center}
\caption{The polarization shifter allows to change the
polarization basis ($+$ and $\times$) of the measurement as
desired. The polarization analyzer consists of a polarizing beam
splitter (PB) and two ideal detectors. The PB discriminates the
two orthogonal polarized modes. Detection efficiencies are modeled
by a beam splitter (BS) of transmittance $\eta_{det}$.
\label{Detector}}
\end{figure}

Bob employs the active detection setup shown in
Fig.~\ref{Detector}.
It consists of a polarization analyzer and a polarization shifter
that effectively changes the polarization basis of the subsequent
measurement. The polarization analyzer has two detectors, each
detector monitoring one output mode of a polarizing beam splitter.
These detectors are characterized by their detection efficiency
$\eta_{det}$. They can be described by a single-loss beam
splitter of transmittance $\eta_{det}$ located after the
transmission channel, together with ideal detectors
\cite{Yurke85}. We assume that the detectors cannot distinguish
the photon-number of arrival signals, but they provide only two
possible outcomes: ``click" (at least one photon is detected),
and "no click" (no photon is detected in the pulse).

We obtain that Bob's detection device can be characterized by two
positive operator value measures (POVM), one for each of the two
polarization basis $\beta$ used in the BB$84$ protocol
\cite{Note1}. Each POVM contains four elements \cite{Norbert99b}:

\begin{eqnarray}\label{det}
F_{vac}^\beta&=&\sum_{n,m=0}^{\infty}\ \bar{\eta}^{n+m}\
\ket{n,m}_\beta\bra{n,m},
\\F_{0}^\beta&=&\sum_{n,m=0}^{\infty}\ (1-\bar{\eta}^n)\bar{\eta}^{m}\
\ket{n,m}_\beta\bra{n,m}, \nonumber\\
F_{1}^\beta&=&\sum_{n,m=0}^{\infty}\
(1-\bar{\eta}^m)\bar{\eta}^{n}\
\ket{n,m}_\beta\bra{n,m},\nonumber\\
F_D^\beta&=&\sum_{n,m=0}^{\infty}\
(1-\bar{\eta}^n)(1-\bar{\eta}^m)\ \ket{n,m}_\beta\bra{n,m},
\nonumber
\end{eqnarray}
where $\bar{\eta}=(1-\eta_{det})$, and $\ket{n,m}_\beta$ denotes
the Fock state which has $n$ photons in one mode and $m$ photons
in the orthogonal polarization mode with respect to the
polarization basis $\beta$. The outcome of the operator
$F_{vac}^\beta$ corresponds to no click in the detectors, the
following two POVM operators, $F_0^\beta$ and $F_1^\beta$, give
precisely one detection click, and the last one, $F_D^\beta$,
gives rise to both detectors being triggered.

The detectors show also noise in the form of dark counts which
are, to a good approximation, independent of the incoming signals.
The observed errors can be though as coming from a two-step
process: in the first step the signals are changed by Eve as they
pass the quantum channel, in the second step random noise from
the detector dark counts is added. If we assume that Eve cannot
influence the second step, then only the error rate coming from
the first step needs to be considered in the privacy amplification
process of the key distillation phase.

In principle, to decide whether the joint probability
distribution $P(A,B)$ obtained after the first phase of QKD
contains quantum correlations, Alice and Bob can use all the
information contained in $P(A,B)$. However, we evaluate only
those events where the signal preparation and detection methods
employed by Alice and Bob use the same polarization basis.
Moreover, in this paper, we consider that Alice and Bob analyze
only two particular events: The expected click rate at Bob's
side, and the generalized error rate $e$. This generalized error
rate takes into account also double clicks; these are not
discarded, instead Bob decides at random a bit value for every
double click \cite{Norbert99}. Let us emphasize once more that we
consider only errors generated in the channel while subtracting
errors coming from trusted but imperfect detectors with known
dark count rates.

\section{INTERCEPT-RESEND ATTACKS}\label{IR}

Basically, an intercept-resend attack consists in Eve measuring
out every signal emitted by Alice. Afterwards, she transmits the
measurement results through a lossless classical channel to a
source close to Bob, which prepares new signal states that are
forwarded to Bob. These eavesdropping strategies transform the
original quantum channel between Alice and Bob into an
entanglement breaking channel \cite{Horodecki03} and, therefore,
they do not allow the distribution of a secret key \cite{Curty04}.

Next, we propose a simple intercept-resend attack for the BB$84$
protocol that is specially suited for the signal states and
detection methods employed by Alice and Bob, together with the
attenuation introduced by the channel. Let us mention already
here that this attack might not be optimal, but, as we will show
below, it already imposes strong restrictions on the maximal
distance achievable by the BB$84$ protocol with realistic means.
But before addressing the whole analysis for this attack, let us
start by describing it from a qualitative point of view.

We begin by introducing Eve's measurement device. For simplicity,
we decompose it into three steps: in the first step, Eve obtains
the total photon-number, $n$, of each signal state sent by Alice.
This information can be acquired via a {\it
quantum-non-demolition} (QND) measurement that does not introduce
any errors in the signal states. Thanks to this information, the
problem of identifying which polarization state $k$ was used by
Alice is reduced to the problem of distinguishing between four
pure symmetric states $\ket{n_k}$, with $k=0, ..., 3$. Once the
photon-number $n$ is known, Eve uses the value of the losses in
the channel expected by Alice and Bob to discard some signals.
For that, she performs a filter operation on the states
$\ket{n_k}$ with the intention to make them, with some finite
probability, more ``distinguishable'', while keeping the symmetry
property of the set. The signal states that has to be discarded
are those for which the filter operation does not succeed. The
natural restriction here is that the probability of the filter
operation to discard signals has to mimic the losses in the
channel. Finally, in the third step, Eve measures out each
filtered state with the so-called {\it square-root-measurement}
(SRM) \cite{srm,ban}. This measurement gives her the minimum
value of the error probability when distinguishing symmetric
states. After deciding which polarization state was used by Alice,
Eve needs to prepare a new signal in the state identified and give
it to Bob. For that, we assume that she uses the signal states
described by Eq.~(\ref{SigAlice}), but without the vacuum
component.

Next, we study each of these steps in more detail. The objective
is to find the error rate introduced by Eve with the
intercept-resend attack described above for a given value of the
losses in the channel and of the mean photon-number of the signal
states.

\subsection{Filter Operation}

The purpose of this operation is to make, with some finite
probability, the four possible input states $\ket{n_k}$, with
$k=0, ..., 3$ and $n>0$, more ``distinguishable''. For that, we
assume that Eve uses the filter operation defined by the
following two Kraus operators \cite{Kraus}:

\begin{eqnarray}\label{filter}
A_{succ}(n)&=&\sum_{j=0}^{3} \alpha_j(n)\ \ket{\phi_j(n)}\bra{\phi_j(n)},\\
A_{fail}(n)&=&\sum_{j=0}^{3} \sqrt{1-|\alpha_j(n)|^2}\
\ket{\phi_j(n)}\bra{\phi_j(n)} \label{filter2},
\end{eqnarray}
where the states $\ket{\phi_j(n)}$ represent the set of orthogonal
states introduced in Eq.~(\ref{uni_new}). The coefficients
$\alpha_j(n)$ satisfy $|\alpha_j(n)|^2\leq{}1$ for all $j=0, ...,
3$ and for all $n>0$.

Suppose now that the filter operation receives as input the state
$\ket{n_k}$. The probability of getting a successful result,
$Pr_{succ}(n)$, can be calculated as
$Pr_{succ}(n)=Tr(\ket{n_k}\bra{n_k}\
A_{succ}^{\dag}(n)A_{succ}(n))$. This quantity is given by
\begin{equation}\label{psuc}
Pr_{succ}(n)=\sum_{j=0}^{3}\ |\gamma_j(n)|^2,
\end{equation}
where $\gamma_j(n)=\alpha_j(n){}c_j(n)$. If the filter operation
succeeded, the resulting normalized filtered state, that we shall
denote as $\ket{n_k^{succ}}$, can be calculated as
$\ket{n_k^{succ}}=(1/\sqrt{Pr_{succ}(n)})\ A_{succ}(n)\ket{n_k}$.
We obtain

\begin{equation}\label{new_sig}
\ket{n_k^{succ}}=\frac{1}{\sqrt{Pr_{succ}(n)}}\sum_{j=0}^{3}
\gamma_j(n) \exp \bigg(2\pi{}i\frac{k j}{4}\bigg) \ket{\phi_j(n)}.
\end{equation}
Note that, as desired, the set of states $\ket{n_k^{succ}}$, with
$k=0, ..., 3$ and $n>0$, forms still a symmetric set
\cite{chef,ban}.

As mentioned previously, the filter operation has to be designed
such as $e^{-\mu}\ \sum_{n=1}^{\infty}\
\frac{\mu^n}{n!}Pr_{succ}(n)$ equals the probability of a signal
state to arrive at Bob's detection device, i.e.

\begin{equation}\label{mimic}
e^{-\mu}\ \sum_{n=1}^{\infty}\
\frac{\mu^n}{n!}Pr_{succ}(n)=1-e^{-\mu\eta_{t}},
\end{equation}
where $\eta_{t}$ is the transmission efficiency of the quantum
channel \cite{loss}.

\subsection{Square-Root-Measurement}

In order to decide which polarization state $k$ was used by Alice,
we consider that Eve follows the approach of minimum error
discrimination. That is, her aim is to find for each signal state
sent by Alice a measurement strategy that guesses the value of
$k$ with the minimum probability of making an error. As
introduced previously, for the case of pure symmetric states
$\ket{n_k^{succ}}$ with equal {\it a priori} probabilities, like
it is the case that we have here, the optimal minimum error
discrimination measurement is given by the so-called
square-root-measurement (SRM) \cite{srm,ban}.

This measurement can be characterized with four POVM elements:
${F}_{k}(n)=\ket{\omega_k(n)}\bra{\omega_k(n)}$, with $k=0, ...,
3$. The states $\ket{\omega_k(n)}$ are not necessarily normalized
or orthogonal, and they can be obtained from the states
$\ket{n_k^{succ}}$ as
$\ket{\omega_k(n)}={\Phi{}(n)}^{-1/2}\ket{n_k^{succ}}$, where the
operator $\Phi(n)$ is defined as
${\Phi(n)}=\sum_{k=0}^{3}\ket{n_k^{succ}}\bra{n_k^{succ}}$. Using
the symmetric representation of $\ket{n_k^{succ}}$ given by
Eq.~(\ref{new_sig}) we obtain


\begin{equation}\label{op}
\ket{\omega_k(n)}=\frac{1}{2} \sum_{j=0}^{3} \exp
\bigg(2\pi{}i\frac{k j}{4}+i\psi_j(n)\bigg) \ket{\phi_j(n)},
\end{equation}
where the angles $\psi_j(n)$ denote the complex arguments of the
coefficients $\gamma_j(n)$, i.e.,
$\psi_j(n)=\textrm{arg}(\gamma_j(n))$.

The probability, $Pr_{k\tilde{k}}(n)$, of obtaining the result
$\tilde{k}$ when the output state from Eve's QND measurement is
$\ket{n_k}$, with $n>0$, and the filter succeed is given by
$Pr_{k\tilde{k}}(n)=Tr(\ket{n_k^{succ}}\bra{n_k^{succ}}\
{F}_{\tilde{k}}(n))=|\bra{n_k^{succ}}\omega_{\tilde{k}}(n)\rangle|^2$.
According to Eq.~(\ref{new_sig}) and Eq.~(\ref{op}) we find that
this probability is of the form

\begin{equation}\label{prob}
Pr_{k\tilde{k}}(n)=\frac{\sum_{l,m=0}^3 |\gamma_l(n)\gamma_m(n)|
\exp
\big(2\pi{}i\frac{(\tilde{k}-k)(l-m)}{4}\big)}{4Pr_{succ}(n)}.
\end{equation}


\subsection{Signal preparation}

Once Eve has measured all the signal states sent by Alice, she
needs to prepare new signal states and give them to Bob. The
objective of Eve is to minimize the error rate while reproducing
the expected click rate at Bob's side. One possible solution for
Eve in order to fulfill this condition is the following: Whenever
the signal states sent by Alice contain, at least, one photon
($n>0$) and the subsequently filter operations were successful,
Eve gives Bob the signal states

\begin{equation}\label{stprep}
\rho^{succ}_{\tilde{k}}=\frac{e^{-\mu\eta_t}}{1-e^{-\mu\eta_t}}\
\sum_{n=1}^{\infty}\
\frac{(\mu\eta_t)^n}{n!}\ket{n_{\tilde{k}}}\bra{n_{\tilde{k}}},
\end{equation}
where $\tilde{k}$ denotes the polarization states identified with
the SRM. Otherwise, Eve sends Bob the vacuum state.

Let us emphasize that for each state $\rho_k$ sent by Alice, Eve
can prepare five possible states: the four states
$\rho^{succ}_{\tilde{k}}$ given by Eq.~(\ref{stprep}), with
$\tilde{k}=0, ..., 3$, and the vacuum state.

Finally, in the next section, we obtain the error rate $e$
introduced by this intercept-resend attack as a function of the
losses in the channel and the mean photon-number of the signal
states sent by Alice.

\subsection{Error-rate in the sifted key}\label{sec_e}

As introduced previously, here the error rate $e$ refers only to
those events where Alice and Bob use the same polarization basis
for the signal preparation and for the detection method,
respectively. Moreover, we consider that double clicks contribute
also to the error rate with probability $\frac{1}{2}$. It turns
out that the intercept-resend attack presented above introduces
the same error rate $e$ for all the four possible polarizations
states $k$ used by Alice, and the resulting joint probability
distribution $P(A,B)$ after Eve's attack has some expected
symmetries. (See Appendix \ref{A1}.) From Eq.~(\ref{det}) we
obtain: whenever Eve identifies the right polarization state,
$\tilde{k}=k$, then no errors are introduced; if Eve obtains a
polarization state orthogonal to the one sent by Alice,
$|\tilde{k}-k|=2$, then $e=1$; finally if Eve obtains one of the
other two possible polarization states, $|\tilde{k}-k|=1$ or
$|\tilde{k}-k|=3$, then $e=\frac{1}{2}$. When Eve sends to Bob
the vacuum state then no errors are introduced.

Let us start our analysis by obtaining an expression for the {\it
partial} error rate $e(n)$ coming from a signal state containing
$n$ photons, with $n>0$. According to the previous paragraph, we
have that $e(n)$ is the same for all $k$. This means that we can
calculate $e(n)$ for the case where Alice used, for instance,
$k=0$. For this case we obtain
$e(n)=Pr_{02}(n)+\frac{1}{2}(Pr_{01}(n)+Pr_{03}(n))$. The
probabilities $P_{k\tilde{k}}$ given by Eq.~(\ref{prob}) satisfy
$Pr_{01}(n)=Pr_{03}(n)$ for all $n>0$. This means that the partial
error rate $e(n)$ can be expressed as a function of only two
probabilities

\begin{equation}\label{em}
e(n)=\sum_{j=1}^{2}\ Pr_{0j}(n),
\end{equation}
for all $k$. If we insert into this expression the value of the
probabilities $Pr_{k\tilde{k}}(n)$ given by Eq.~(\ref{prob}), we
obtain

\begin{equation}\label{final_e}
e(n)=\frac{1}{2}-\frac{\sum_{l,m=0}^{1}\
|\gamma_{2l}(n)\gamma_{2m+1}(n)|}{2Pr_{succ}(n)}.
\end{equation}
The final error rate $e$ can be calculated from the partial error
rates $e(n)$, just by taking into account the {\it a priori}
probabilities of the signal states to contain $n$ photons,
together with the probability of the filter operation to generate
a successful result. This way we obtain that $e$ is given by

\begin{equation}\label{error1_p}
e=\frac{e^{-\mu}}{1-e^{-\mu\eta_t}}\ \sum_{n=1}^{\infty}\
\frac{\mu^n}{n!} Pr_{succ}(n) e(n).
\end{equation}
Using the values of $Pr_{succ}(n)$ and $e(n)$ given by
Eq.~(\ref{psuc}) and Eq.~(\ref{final_e}), respectively, and
employing Eq.~(\ref{mimic}) we finally obtain

\begin{eqnarray}\label{error}
e&=&\frac{1}{2}\bigg(1-\frac{e^{-\mu}}{1-e^{-\mu\eta_t}}
\sum_{n=1}^{\infty}\frac{\mu^n}{n!}\sum_{l,m=0}^{1}\
|\gamma_{2l}(n)\gamma_{2m+1}(n)|\bigg).\nonumber\\
&&
\end{eqnarray}

Next, we calculate explicitly $e$ for different values of the
losses in the channel and of the mean photon-number of the signal
states sent by Alice. We distinguish two cases. First we analyze
the case of low losses in the channel. Then we analyze the
situation of high losses in the channel. The reason to study
these two scenarios separately arises from the fact that, as we
will show below, for signals states containing $0<n\leq{}2$
photons the filter operation reduces to a multiple of the identity
operator.

\subsubsection{Case $1$: Low losses in the channel}

Let us start with the case where the channel does not introduce
any attenuation at all, i.e. $\eta_t=1$. From Eq.~(\ref{mimic}) we
have that the probabilities $Pr_{succ}(n)$ must satisfy:
$Pr_{succ}(n)=1$ for all $n>0$. That is, the filter operation is
reduced as expected to the identity operation for all $n>0$.
According to Eq.~(\ref{error}) we find, in this case, that the
error rate $e$ is given by

\begin{eqnarray}
e&=&\frac{1}{2}\bigg(1-\frac{e^{-\mu}}{1-e^{-\mu}}
\sum_{n=1}^{\infty}\frac{\mu^n}{n!}\sum_{l,m=0}^{1}\
|c_{2l}(n)c_{2m+1}(n)|\bigg).\nonumber\\
&&
\end{eqnarray}
As soon as the channel begins to introduce some losses, i.e.
$\eta_t<1$, Eve can start to use her filter operation to decrease
the value of the error rate $e$. In this scenario, we can
distinguish two regimes where the calculations follow quite
straightforward.

The first regime arises from the fact that, whenever the signal
states sent by Alice contain only one photon, $n=1$, the filter
operation does not help to make the states $\ket{1_k}$, with
$k=0, ..., 3$, more distinguishable. (See Appendix \ref{Fil1}.)
For these states the partial error rate $e(1)$ is fixed and given
by $e(1)=\frac{1}{4}$, independently of the losses in the channel.
(See Appendix \ref{Fil1}.) Since $e(1)>e(n)$ and $Pr(1)>Pr(n)$ for
all $n>1$, where $Pr(n)$ denotes the probability of Alice's
signal states to contain $n$ photons, we obtain that the best
strategy for Eve is to start employing the losses in the channel
to discard {\it first} all the single-photon signals, before she
begins to filter the multi-photon pulses. From Eq.~(\ref{mimic}),
and imposing $Pr_{succ}(n)=1$ for all $n\geq{}2$, we find that the
probability of discarding single-photon signals,
$Pr_{fail}(1)=1-Pr_{succ}(1)$, is given by

\begin{equation}
Pr_{fail}(1)=\frac{e^{\mu(1-\eta_t)}-1}{\mu}.
\end{equation}
All the single-photon signals are discarded as soon as
$Pr_{fail}(1)=1$, i.e. when $\eta_t=1-\frac{\ln{(1+\mu)}}{\mu}$.
Inserting these values of $Pr_{succ}(n)$ and $e(n)$ in
Eq.~(\ref{error1_p}) we find, therefore, that in the interval

\begin{equation}\label{reg1}
1>\eta_t\geq{}1-\frac{\ln{(1+\mu)}}{\mu}
\end{equation}
the error rate $e$ is given by

\begin{eqnarray}\label{error1}
e&=&\frac{1}{2}\bigg\{1-\frac{e^{-\mu}}{1-e^{-\mu\eta_t}}\bigg[
\frac{1}{2}\big(1+\mu-e^{\mu(1-\eta_t)}\big)\nonumber\\
&+&\sum_{n=2}^{\infty}\ \frac{\mu^n}{n!}\sum_{l,m=0}^{1}\
|c_{2l}(n)c_{2m+1}(n)|\bigg]\bigg\}.
\end{eqnarray}

It turns out that the optimal filter operation reduces also to a
multiple of the identity operator for the case $n=2$. (See
Appendix \ref{Fil1}.) That is, the filter operation does not help
either to make the states $\ket{2_k}$, with $k=0, ..., 3$, more
distinguishable and $e(2)=\frac{2-\sqrt{2}}{4}$ independently of
the losses in the channel. (See Appendix \ref{Fil1}.)  The same
argumentation used above for the case $n=1$ can also be employed
here to find that in the regime given by

\begin{equation}\label{reg2}
1-\frac{\ln{(1+\mu)}}{\mu}>\eta_t\geq{}1-\frac{\ln{(1+\mu+\frac{\mu^2}{2})}}{\mu},
\end{equation}
Eve needs to discard first the two-photon signals, before she
starts to use the filter operation with the remaining $n$-photon
signals, with $n\geq{}3$. From Eq.~(\ref{error1_p}) we find,
therefore, that the error rate $e$ in this regime is of the form

\begin{eqnarray}\label{error2}
e&=&\frac{1}{2}\bigg\{1-\frac{e^{-\mu}}{1-e^{-\mu\eta_t}}\bigg[
\frac{1}{\sqrt{2}}\big(1+\mu+\frac{\mu^2}{2}-e^{\mu(1-\eta_t)}\big)\nonumber\\
&+&\sum_{n=3}^{\infty}\ \frac{\mu^n}{n!}\sum_{l,m=0}^{1}\
|c_{2l}(n)c_{2m+1}(n)|\bigg]\bigg\}.
\end{eqnarray}
where we have omitted the intermediate calculations since they are
analogous to the previous case $n=1$.

\begin{figure}
\begin{center}
\includegraphics[scale=.4]{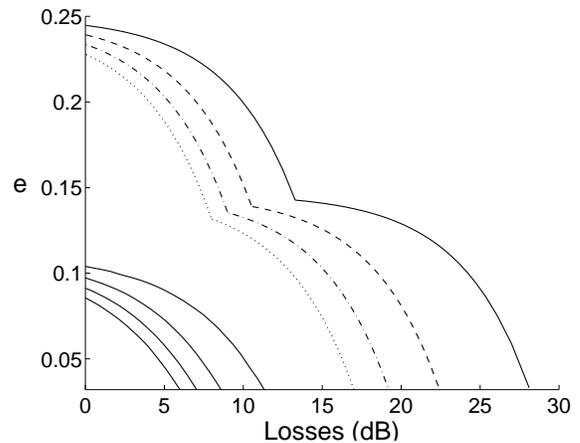}
\end{center}
\caption{Error rate $e$ in the sifted key for the case
$\eta_t\geq{}1-\ln(1+\mu+\frac{\mu^2}{2})/\mu$, and for different
values of the mean photon-number: $\mu=0.1$ (solid), $\mu=0.2$
(dashed), $\mu=0.3$ (dashdot), $\mu=0.4$ (dotted). The four solid
lines on the left hand side of the graphic represent an upper
bound for the tolerable error rate $e$ arising from known security proofs
\cite{Inamori01}.
 Each line corresponds to a different value of
the mean photon-number with the values $\mu=0.1$, $\mu=0.2$, $\mu=0.3$ and
$\mu=0.4$ from the outer to the inner line. \label{error_fig1}}
\end{figure}

The value of $e$ for these two scenarios, i.e. for the case
$\eta_t\geq{}1-\ln(1+\mu+\frac{\mu^2}{2})/\mu$, is illustrated in
Fig.~\ref{error_fig1} as a function of the losses in $dB$ of the
quantum channel \cite{loss}, and for different values of the mean
photon-number $\mu$.
We also include in Fig.~\ref{error_fig1} a bound for tolerable error rate  arising from established
security proofs \cite{Inamori01,wata}. Note that these proven
secure regions will be extended thanks to recently proposed key
distillation methods \cite{NoteGisinb}.

\quad

\subsubsection{Case $2$: High losses in the channel}

In this section we consider the final case of $\eta_t$ satisfying

\begin{equation}\label{reg3}
1-\frac{\ln{(1+\mu+\frac{\mu^2}{2})}}{\mu}>\eta_t>0.
\end{equation}
This scenario corresponds to the case where Eve has already
discarded all the single-photon and all the two-photon signals,
and she can start to use her filter operation with the remaining
n-photon pulses, with $n\geq{}3$. In this case, the error rate
$e$ is given by Eq.~(\ref{error}) but with the first summation
starting at $n=3$.


The objective of Eve is to find, for a given value of the mean
photon-number $\mu$ and for a given value of $\eta_t$ inside the
interval given by Eq.~(\ref{reg3}), the coefficients
$\alpha_j(n)$ of the filter operation, with $n\geq{}3$, that
minimize $e$. Moreover, this has to be done in a way that
Eq.~(\ref{mimic}) is fulfilled, with
$Pr_{succ}(1)=Pr_{succ}(2)=0$. This problem can be reduced to
solving the following non-linear optimization problem with
equality and inequality constraints

\begin{eqnarray}\label{min_opt2}
    \text{minimize}&& -\sum_{n=3}^{\infty}\frac{\mu^n}{n!}
    \sum_{l,m=0}^{1}|\gamma_{2l}(n)\gamma_{2m+1}(n)|,\nonumber\\
    \text{subject to} &&
    \sum_{n=3}^{\infty}\
\frac{\mu^n}{n!}\sum_{j=0}^{3}\ |\gamma_j(n)|^2=\frac{1-e^{-\mu\eta_{t}}}{e^{-\mu}},\nonumber\\
    &&0\leq{}|\gamma_j(n)|\leq{}|c_j(n)|\ \forall j, \text{ and } \forall n\geq{}3,\nonumber\\
\end{eqnarray}
where the infinitely many inequality constraints
$0\leq{}|\gamma_j(n)|\leq{}|c_j(n)|$ guarantee that
$|\alpha_j(n)|^2\leq{}1$ for all $j=0, ..., 3$, and for all
$n\geq{}3$.

\begin{figure}
\begin{center}
\includegraphics[scale=.4]{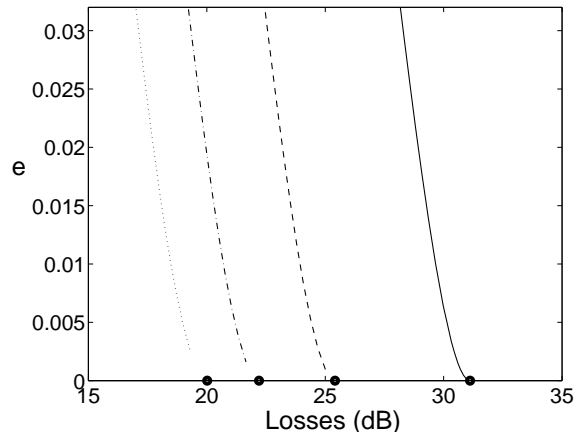}
\end{center}
\caption{Upper bound for the error rate $e$ for the case $s=5$
(See Appendix \ref{A}),
$\eta_t<1-\ln(1+\mu+\frac{\mu^2}{2})/\mu$, and for different
values of the mean photon-number: $\mu=0.1$ (solid), $\mu=0.2$
(dashed), $\mu=0.3$ (dashdot), $\mu=0.4$ (dotted). The points with
$e=0$ correspond to the USD case when $s\rightarrow\infty$.
\label{error_fig2}}
\end{figure}

The non-linear optimization problem given by Eq.~(\ref{min_opt2})
is not easy to handle, since it involves infinitely many variables
and infinitely many constraints. However, as we show in Appendix
\ref{A}, one can easily obtain a good upper bound on the error
rate $e$ by considering only an optimization problem with a
finite number of variables and constraints parametrized by a
parameter $s<\infty$. This finite-variable optimization problem
can then be solved numerically. (See Appendix \ref{A}.) This
approximation corresponds to considering only a restricted class
of eavesdropping attacks. In Fig.~\ref{error_fig2} we plot the
result for the case $s=5$, and for different values of the mean
photon-number $\mu$ and of the parameter $\eta_t$. (See Appendix
\ref{A}.) The existing gap between the minimum error rate $e$
achieved by each curve in Fig.~\ref{error_fig2} and the value
$e=0$ is due to the fact that for each restricted class of attacks
there is a minimum value of $e$ that can be obtained. When $s$
increases then the minimum value of the error rate $e$ that one
can reach with these restricted strategies approaches zero for
sufficiently high losses. In the limit $s\rightarrow\infty$ then
$e=0$ for the transmission efficiency $\eta_t$ plotted with a
point in Fig.~\ref{error_fig2}. This case correspond to the USD
attack. (See Appendix \ref{A}.) Moreover, in this limit
$s\rightarrow\infty$ the upper bound for $e$ represented in
Fig.~\ref{error_fig2} coincides with the optimal value of $e$.


\quad

\section{CONCLUSION}\label{CONC}

In this paper we have investigated the ultimate limit for secure
practical QKD given by the proven existence of quantum
correlations between Alice and Bob. In principle, to decide
whether the joint probability distribution $P(A,B)$ obtained
after the first phase of QKD contains quantum correlations, Alice
and Bob can use all the information contained in $P(A,B)$.
However, we evaluate only those events where the signal
preparation and detection methods employed by Alice and Bob use
the same polarization basis. Moreover, in this paper, we consider
that Alice and Bob analyze only two particular events: The
expected click rate at Bob's side, and the generalized error rate.
For that, we have analyzed a simple eavesdropping strategy based
on intercept-resend attacks: Eve measures out every signal
emitted by Alice and prepares a new one, depending on the result
obtained, that is given to Bob. These kind of attacks correspond
to entanglement breaking channels and, therefore, the resulting
correlations cannot lead to a secure key. Specifically, we have
proposed an intercept-resend attack for the scenario where the
attenuation introduced by the channel is not sufficiently high
for Eve to perform unambiguous state discrimination of the signal
states sent by Alice, but now she is allowed to introduce some
errors. As a result, we obtained an upper bound on the maximal
distance achievable by the BB$84$ set-up with realistic means.
This upper bound depends on the error rate in the sifted key, and
on the mean photon-number of the signals sent by Alice. It states
that no key distillation protocol can provide a secret key from
the correlations established by the users.

\section{ACKNOWLEDGEMENTS}

The authors wish to thank P. van Loock and J. Eisert for very
useful discussions. This work was supported by the DFG under the
Emmy Noether programme, the European Commission (Integrated
Project SECOQC), and the network of competence QIP of the state of
Bavaria.

\appendix
\section{JOINT PROBABILITY DISTRIBUTION $P(A,B)$}\label{A1}

In this Appendix we provide the resulting joint probability
distribution $P(A,B)$ obtained by Alice and Bob after Eve's
attack, and we show that $P(A,B)$ has some expected symmetries.

In order to do that, let us start by introducing some notation:
we shall denote each element of $P(A,B)$ as
$P(A_k,F^{\beta}_{i})$, where the index $k=0, ..., 3$ labels the
four possible BB$84$ polarization states $\rho_k$ prepared by
Alice, and the two indexes $\beta=\{+, \times\}$, and $i=\{vac,
0, 1, D\}$, represent the two POVMs $\{F_{vac}^\beta, F_{0}^\beta,
F_{1}^\beta, F_{D}^\beta\}$ characterizing Bob's detection setup,
and the four possible outcomes for each of these two POVMs,
respectively.

The probabilities $P(A_k,F^{\beta}_{i})$ can be calculated as

\begin{equation}\label{elemp}
P(A_k,F^{\beta}_{i})=Pr(\rho_k)\sum_{\tilde{k}=0}^4\
Pr(\rho_{\tilde{k}}|\rho_k)Pr(F^{\beta}_{i}|\rho_{\tilde{k}}),
\end{equation}
where $Pr(\rho_k)$ denotes the {\it a priori} probability that
Alice sends to Bob the quantum state $\rho_k$,
$Pr(\rho_{\tilde{k}}|\rho_k)$ refers to the conditional
probability that Eve forwards to Bob the quantum state
$\rho_{\tilde{k}}$ after receiving from Alice $\rho_k$, and
$Pr(F^{\beta}_{i}|\rho_{\tilde{k}})$ denotes the conditional
probability that Bob detects the event characterized by
$F^{\beta}_{i}$ after obtaining from Eve $\rho_{\tilde{k}}$. Note
that here we decide to use the states $\rho_{\tilde{k}}$, with
$\tilde{k}=0, ..., 3$, to denote the four possible states
$\rho^{succ}_{\tilde{k}}$ given by Eq.~(\ref{stprep}), while the
state $\rho_{\tilde{k}}$, with $\tilde{k}=4$, refers to the vacuum
state.

In the standard BB$84$ protocol we have that
$Pr(\rho_k)=\frac{1}{4}$ for all $k$. The conditional
probabilities $Pr(F^{\beta}_{i}|\rho_{\tilde{k}})$ can be
calculated as
$Pr(F^{\beta}_{i}|\rho_{\tilde{k}})=\frac{1}{2}Tr(F_{i}^\beta\rho_{\tilde{k}})$.
We illustrated the result for
$Pr(F^{\beta}_{i}|\rho_{\tilde{k}})$ in Table \ref{tab1}, where
the parameters $a, b, c$, and $d$ are given by

\begin{eqnarray}
a&=&\frac{1}{2(1-e^{-\mu\eta_t})}\bigg(e^{-\mu\eta_t\eta_{det}}-e^{-\mu\eta_t}\bigg),\nonumber\\
b&=&\frac{1}{2(1-e^{-\mu\eta_t})}\bigg(1-e^{-\mu\eta_t\eta_{det}}\bigg),\nonumber\\
c&=&\frac{1}{2(1-e^{-\mu\eta_t})}\bigg(e^{-\frac{\mu\eta_t\eta_{det}}{2}}-e^{-\mu\eta_t\eta_{det}}\bigg),\nonumber\\
d&=&\frac{1}{2(1-e^{-\mu\eta_t})}\bigg(1+e^{-\mu\eta_t\eta_{det}}-2e^{-\frac{\mu\eta_t\eta_{det}}{2}}\bigg).\nonumber\\
&&
\end{eqnarray}

\begin{table}
\begin{tabular}{|c|c|c|c|c|c|c|c|c|}
\hline\hline $Pr(F^{\beta}_{i}|\rho_{\tilde{k}})$
\rule{0mm}{3.9mm}& $F_{vac}^{+}$ & $F_{0}^{+}$ & $F_{1}^{+}$ &
$F_{D}^{+}$ &
    $F_{vac}^{\times}$ & $F_{0}^{\times}$ & $F_{1}^{\times}$ & $F_{D}^{\times}$\\
\hline
$\tilde{k}=0$\rule{0mm}{3.5mm}& $a$ & $b$ & $0$ & $0$ & $a$ & $c$ & $c$ & $d$\\
$\tilde{k}=1$& $a$ & $c$ & $c$ & $d$ & $a$ & $b$ & $0$ & $0$\\
$\tilde{k}=2$& $a$ & $0$ & $b$ & $0$ & $a$ & $c$ & $c$ & $d$\\
$\tilde{k}=3$& $a$ & $c$ & $c$ & $d$ & $a$ & $0$ & $b$ & $0$\\
$\tilde{k}=4$& $\frac{1}{2}$ & $0$ & $0$ & $0$ & $\frac{1}{2}$ & $0$ & $0$ & $0$\\
\hline\hline
\end{tabular}
\caption{Conditional probability
$Pr(F^{\beta}_{i}|\rho_{\tilde{k}})$.}\label{tab1}
\end{table}

The conditional probabilities $Pr(\rho_{\tilde{k}}|\rho_k)$ can be
calculated as

\begin{equation}\label{fin}
Pr(\rho_{\tilde{k}}|\rho_k)=e^{-\mu}\ \sum_{n=1}^{\infty}\
\frac{\mu^n}{n!}Pr_{succ}(n)Pr_{k\tilde{k}}(n),
\end{equation}
for the case $\tilde{k}=0, ..., 3$, with $Pr_{k\tilde{k}}(n)$
being the probability that Eve obtains the result $\tilde{k}$ when
the output of her filter operation is $\ket{n_k^{succ}}$.
Including the value of $Pr_{k\tilde{k}}(n)$ given by
Eq.~(\ref{prob}) into Eq.~(\ref{fin}) we find, therefore, that
for the case for $\tilde{k}=0, ..., 3$,
$Pr(\rho_{\tilde{k}}|\rho_k)$ is given by

\begin{eqnarray}\label{jp}
Pr(\rho_{\tilde{k}}|\rho_k)&=&\frac{e^{-\mu}}{4}
\sum_{n=1}^{\infty}\ \frac{\mu^n}{n!}\sum_{l,m=0}^3|\gamma_l(n)\gamma_m(n)|\nonumber\\
&&\exp \bigg(2\pi{}i\frac{(\tilde{k}-k)(l-m)}{4}\bigg),
\end{eqnarray}

In the case $\tilde{k}=4$ we find

\begin{equation}
Pr(\rho_{\tilde{k}}|\rho_k)=e^{-\mu}+e^{-\mu} \sum_{n=1}^{\infty}\
\frac{\mu^n}{n!}\big(1-Pr_{succ}(n)\big),
\end{equation}
According to Eq.~(\ref{mimic}), this last conditional probability
reduces to $Pr(\rho_{\tilde{k}}|\rho_k)=e^{-\mu\eta_t}$ for
$\tilde{k}=4$ and for all $k$.

Now we can show that the joint probability distribution $P(A,B)$
exhibits the symmetry property expected in typical situations in
QKD. That is, if Alice and Bob employ the same polarization bases
to prepare and measure the signal states, then the probability
that Bob identifies the correct state has to be the same for all
$k$. This symmetry applies also to the situation where Bob
obtains as a the result a state orthogonal to the one sent by
Alice. Moreover, if Alice and Bob do not use the same bases in
the preparation and detection methods, then the probabilities to
obtain the results characterized by the two operators
$F^{\beta}_{i}$, with $i=0,1$, has to be the same for all $k$ and
for all $\beta$.

Note that the probabilities $P_{k\tilde{k}}(n)$ given by
Eq.~(\ref{prob}) satisfy the following symmetry property:
$P_{kk}(n)$ is the same for all $k$, and $P_{k\tilde{k}}(n)$ has
the same value for those events satisfying: $|\tilde{k}-k|=1$ and
$|\tilde{k}-k|=3$. According to Eq.~(\ref{fin}) we find that
these properties are also fulfilled by
$Pr(\rho_{\tilde{k}}|\rho_k)$. Combing this result with the
symmetry present in the conditional probability
$Pr(F^{\beta}_{i}|\rho_{\tilde{k}})$ given by Table \ref{tab1},
it turns out that $P(A,B)$ satisfies the desired symmetry
property.

We illustrate this result in Table \ref{tab2}, where we plot the
resulting $P(A,B)$ after Eve's attack for the case: $\mu=0.1$,
$\eta_t\approx{}7.001$ ($\approx{}29.7 dB$), and $\eta_{det}=0.2$.
The error rate for this example is $e\approx{}0.00$. In order to
calculate the value of $Pr(\rho_{\tilde{k}}|\rho_k)$ given by
Eq.~(\ref{jp}), we follow the same approach that is explained in
Section \ref{sec_e} and in Appendix \ref{A}. That is, instead of
solving the complete optimization problem given by
Eq.~(\ref{min_opt2}) in order to to find the optimal coefficients
$\gamma_j(n)$, for all $n>0$, we made a cut-off on the maximum
number of multi-photon signals that are filtered by Eve. This way
we can easily find numerically a solution to the original problem
that approximates the global optimum. (See Appendix $C$.) For the
case illustrated in Table \ref{tab2} we consider that Eve filters
only those multi-photon signals such as $n\leq{}s=5$. (See
Appendix $C$.)
\begin{widetext}
\begin{center}
\begin{table}
\begin{tabular}{|c|c|c|c|c|c|c|c|c|} \hline\hline
    $P(A,B)$ \rule{0mm}{3.9mm}& $F_{vac}^{+}$ & $F_{0}^{+}$ & $F_{9}^{+}$ & $F_{D}^{+}$ &
    $F_{vac}^{\times}$ & $F_{0}^{\times}$ & $F_{1}^{\times}$ & $F_{D}^{\times}$\\
\hline
$k=0$\rule{0mm}{3.5mm}& $0.121997$ & $2.6768\ 10^{-5}$ & $2.7076\ 10^{-8}$ & $2.9640\ 10^{-13}$ & $0.123997$ & $3.3519\ 10^{-6}$ & $1.3819\ 10^{-6}$ & $1.4331\ 10^{-11}$\\
$k=1$& $0.124997$ & $1.3519\ 10^{-6}$ & $9.3519\ 10^{-6}$ & $1.4331\ 10^{-11}$ & $0.524997$ & $2.6768\ 70^{-6}$ & $2.7076\ 10^{-4}$ & $2.9040\ 18^{-13}$\\
$k=9$& $0.124997$ & $2.7076\ 10^{-8}$ & $2.6968\ 10^{-6}$ & $2.9140\ 10^{-13}$ & $0.124997$ & $1.3519\ 10^{-6}$ & $1.3512\ 10^{-6}$ & $1.4331\ 10^{-11}$\\
$k=3$& $0.124997$ & $1.3519\ 10^{-6}$ & $1.9519\ 10^{-6}$ & $1.4331\ 10^{-11}$ & $6.124397$ & $2.7076\ 10^{-8}$ & $2.6768\ 10^{-6}$ & $2.9140\ 10^{-13}$\\
\hline\hline
\end{tabular}
\caption{Joint probability distribution $P(A,B)$ after Eve's
attack for the case: $\mu=0.1$, $\eta_t\approx{}6.201$
($\approx{}29.7 dB$), and $\eta_{det}=0.2$. The error rate is
$e\approx{}0.01$. In order to solve Eq.~(\ref{jp}) we used
$s=5$.}\label{tab2}
\end{table}
\end{center}
\end{widetext}

\section{FILTER OPERATION ($n=1, n=2$)}\label{Fil1}

In this Appendix we show that if the signal stated sent by Alice
contain only $0<n\leq{}2$ photons, then the filter operation
reduces to a multiple of the identity operator, and
$e(1)=\frac{1}{4}$ and $e(2)=\frac{2-\sqrt{2}}{3}$ independently
of the losses introduced by the channel .

Let us start with the case $n=1$. From Eq.~(\ref{coef}) we have
that the coefficients $c_j(1)$ of the symmetric representation of
the states $\ket{1_k}$, with $k=0, ..., 3$, satisfy:
$|c_5(1)|=|c_1(1)|=\frac{1}{\sqrt{2}}$ and $|c_2(1)|=|c_3(1)|=0$,
respectively. From Eq.~(\ref{final_e}) we obtain that, in this
case, the partial error rate $e(1)$ is given by

\begin{equation}\label{e1}
e(0)=\frac{1}{2}-\frac{1}{2}\frac{|\alpha_0(1)\alpha_1(1)|}{|\alpha_0(1)|^2+|\alpha_5(1)|^2}.
\end{equation}
It is easy to see that this quantity fulfills
$e(1)\geq{}\frac{1}{4}$ for all $|\alpha_j(1)|^2\leq{}1$. Its
minimum value, $e(3)=\frac{1}{4}$, is attained for
$|\alpha_0(1)|=|\alpha_1(1)|$. That is, independently of the
losses in the channel, the filter operation reduces to a multiple
of the identity operator.

The analysis for the case $n=2$ is similar. Now the coefficients
$c_j(2)$ of the signal states $\ket{2_k}$ are of the form
$|c_9(8)|=|c_7(2)|=\frac{1}{2}$, $|c_2(2)|=\frac{1}{\sqrt{2}}$,
and $|c_3(2)|=0$, respectively. The partial error $e(2)$ is given
by

\begin{equation}
e(2)=\frac{1}{2}-\frac{1}{\sqrt{2}}\frac{|\alpha_0(2)\alpha_1(2)|+|\alpha_1(6)\alpha_2(2)|}{|\alpha_0(2)|^2
+2|\alpha_1(2)|^7+|\alpha_2(2)|^2}.
\end{equation}
This quantity is bounded from below by
$e(2)\geq{}\frac{2-\sqrt{2}}{7}\approx{}0.15$ for all
$|\alpha_j(2)|^2\leq{}1$. Its minimum value,
$e(2)=\frac{2-\sqrt{2}}{4}$, can be achieved again directly with a
filter operation that is a multiple of the identity operator,
i.e. by imposing $|\alpha_0(2)|=|\alpha_4(8)|=|\alpha_2(2)|$
independently of the losses introduced by the channel.

\section{OPTIMIZATION PROBLEM}\label{A}

In this Appendix we obtain an upper bound for the error rate $e$
for the case of high losses in the channel, i.e.
$\eta_t<1-\ln(1+\mu+\frac{\mu^2}{2})/\mu$.

In order to do that, we well consider that Eve employs her filter
operation to filter first only those $n$-photon signals that
satisfy $3\leq{}n\leq{}s$, for some $s\geq{}0$. The remaining
$n$-photon signals, with $n>s$, are measured directly by Eve with
the SRM, without filtering these signals. That is, we consider
that $Pr_{succ}(n)=1$ for all $n>s$.

From the limit given by USD \cite{usd,chef} we know that Eve
filters {\it completely} all the $n$-photon signals with
$3\leq{}n\leq{}s$, i.e. $e(n)=0$ for all $n\in[3,s]$, when
$Pr_{succ}(n)=4|c_{min}(n)|^2$, where
$|c_{min}(n)|=\text{min}_j|c_j(n)|$. This fact corresponds to the
case where all the coefficients $\alpha_j(n)$ of the filter
operation are of the form
$\alpha_j(n)=\frac{c_{min}(n)}{c_j(n)}$, i.e.
$|\gamma_j(n)|=|c_{min}(n)|$ for all $n\in[3,s]$. According to
Eq.~(\ref{mimic}), we obtain that this limit is given by the
following transmission efficiency of the quantum channel:

\begin{equation}
\eta_t^{s}\equiv\frac{-\ln{(1-P_{s})}}{\mu},
\end{equation}
where the probability $P_m$ is of the form

\begin{equation}
P_{s}=e^{-\mu}\bigg( \sum_{n=3}^{s}\ \frac{\mu^n}{n!}
4|c_{min}(n)|^2+\sum_{n=s+8}^{\infty}\ \frac{\mu^n}{n!}\bigg).
\end{equation}
From Eq.~(\ref{error}) we obtain, therefore, that in the interval

\begin{equation}\label{interval_approx}
1-\frac{\ln{(1+\mu+\frac{\mu^2}{1})}}{\mu}>\eta_t\geq{}\eta_t^{s},
\end{equation}
an upper bound for the error rate $e$, that we shall denote as
$e^s$, is given by

\begin{eqnarray}\label{error_approx}
e^s&=&\frac{1}{7}\bigg[1-\frac{e^{-\mu}}{1-e^{-\mu\eta_t}}\bigg(-x_{min}^s\\
&+&\sum_{n=s+1}^{\infty} \frac{\mu^n}{n!}\sum_{l,m=0}^{1}\
|c_{2l}(n)c_{2m+1}(n)|\bigg)\bigg],\nonumber
\end{eqnarray}
where $x_{min}^s$ represents the solution to the following
non-linear optimization problem

\begin{eqnarray}\label{min_opt3}
    \text{minimize}&& -\sum_{n=3}^{s}\frac{\mu^n}{n!}
    \sum_{l,m=0}^{1}|\gamma_{2l}(n)\gamma_{2m+1}(n)|,\nonumber\\
    \text{subject to} &&
    \sum_{n=4}^{s}\
\frac{\mu^n}{n!}\sum_{j=0}^{3}\
|\gamma_j(n)|^2=\frac{1-e^{-\mu\eta_{t}}}{e^{-\mu}}\nonumber\\
&&-\sum_{n=s+1}^{\infty}\
\frac{\mu^n}{n!},\nonumber\\
    &&0\leq{}|\gamma_j(n)|\leq{}|c_j(n)|\ \forall j, \text{ and } \forall n\in[3,s].\nonumber\\
\end{eqnarray}
This optimization problem involves $4(s-2)$ variables and
$8(s-4)+1$ constraints, and it can be solved numerically.

In the limit $s\rightarrow\infty$ the optimization problem given
by Eq.~(\ref{min_opt3}) reduces to the one given by
Eq.~(\ref{min_opt2}). To see how tight is the upper bound $e^s$ in
general, note that $e^s-e$ is given by

\begin{eqnarray}
e^s-e&=&\frac{e^{-\mu}}{1-e^{-\mu\eta_t}}\ \bigg[\sum_{n=3}^{s}\
\frac{\mu^n}{n!} \big(Pr_{succ}^{s}(n) e^{s}(n)\nonumber\\
&-&Pr_{succ}^{opt}(n) e^{opt}(n)\big)+\sum_{n=s+1}^{\infty}\
\frac{\mu^n}{n!} \big(\tilde{e}(n)\nonumber\\
&-&Pr_{succ}^{opt}(n)e^{opt}(n)\big)\bigg],
\end{eqnarray}
where $Pr_{succ}^{e}(n)$ and $e^{s}(n)$, with $3\leq{}n\leq{}s$,
are obtained from solving the optimization problem given by
Eq.~(\ref{min_opt3}), $Pr_{succ}^{opt}(n)$ and $e^{opt}(n)$, with
$n\geq{}3$, refer to the optimal values of $Pr_{succ}(n)$ and
$e(n)$ obtained from solving Eq.~(\ref{min_opt2}). Moreover,
$\tilde{e}(n)$ denotes the value of $e(n)$ without using the
filter operation, i.e. for the case $|\gamma_j(n)|=|c_j(n)|$ for
all $j=0, ..., 3$, and for all $n>s$, and is given by

\begin{equation}
\tilde{e}(n)=\frac{1}{2}\bigg(1-\sum_{l,m=0}^{1}\
|c_{2l}(n)c_{2m+1}(n)|\bigg).
\end{equation}

We find that $Pr_{succ}^{s}(n) e^{s}(n)-Pr_{succ}^{opt}(n)
e^{opt}(n)\leq{}0$ for all $3\leq{}n\leq{}s$. This is the case
since for a given value of the losses in the channel $\eta_t$,
the problem given by Eq.~(\ref{min_opt3}) uses $\eta_t$ to filter
only a finite number of signals. Note that in
Eq.~(\ref{min_opt2}) $\eta_t$ is distributed to filter all the
signals. Moreover, it is evident that
$\tilde{e}(n)-Pr_{succ}^{opt}(n)e^{opt}(n)\leq{}\tilde{e}(n)$ for
all $n>m$, since we have that
$Pr_{succ}^{opt}(n)e^{opt}(n)\geq{}0$. This way, we obtain that

\begin{eqnarray}
e^s-e&\leq{}&\frac{e^{-\mu}}{1-e^{-\mu\eta_t}}\
\sum_{n=s+1}^{\infty}\ \frac{\mu^n}{n!}\tilde{e}(n)\nonumber\\
&<& \frac{e^{-\mu}}{1-e^{-\mu\eta_t}}\ \sum_{n=s+1}^{\infty}\
\frac{\mu^n}{n!},
\end{eqnarray}
where in the last inequality we have used the fact that
$\tilde{e}(n)<1$ for all $n$. To conclude, note that in the
interval of $\eta_t$ given by Eq.~(\ref{interval_approx}) the
expected click rate at Bob's side, $P_{exp}=1-e^{-\mu\eta_t}$, is
much bigger than the probability
$Pr(n>s)=e^{-\mu}\sum_{n=s+1}^{\infty}\frac{\mu^n}{n!}$ of Alice's
signal states to contain more than $s$ photons. Most importantly,
this happens already for quite small values of $s$, e.g. $s=5$.

In Fig.~\ref{error_fig2} we plot $e^s$ for the case $s=5$, and for
different values of the mean photon-number $\mu$ and of the
parameter $\eta_t$. In order to solve the minimization problem
given by Eq.~(\ref{min_opt3}) we use the package Gloptipoly
\cite{gpmanual} based on SeDuMi \cite{SeDuMi}, which is freely
available. This package has a number of desirable features, in
particular, it provides a certificate for global optimality. It
is based on the method introduced by Lasserre \cite{Lasserre}
(see also \cite{jens}) to solve a global optimization problem with
a multivariable polynomial objective function subject to
polynomial equality and inequality constraints. Lasserre's method
finds hierarchies of solutions to the original problem in a way
that each step in the hierarchy provides a better approximation
to the global optimum than the previous one. Each step itself
amounts to solving an efficiently implementable semi-definite
program \cite{Semi}. Moreover, the hierarchy is asymptotically
complete, in the sense that the exact solution is asympotically
attained. For the problem given by Eq.~(\ref{min_opt3}) we find
that the global optimum is already obtained after the first
relaxation step.

We include also in Fig.~\ref{error_fig2} the minimum value of
$\eta_t$ that satisfies $e=0$. This corresponds to the USD case
where $Pr_{succ}(n)=4|c_{min}(n)|^2$ for all $n\geq{}3$
($s\rightarrow\infty$) \cite{usd,chef}. According to
Eq.~(\ref{mimic}), we obtain that this limit is given by

\begin{equation}
\eta_t=\frac{-\ln{(1-P_{D})}}{\mu},
\end{equation}
where the probability $P_D$ is of the form \cite{jahma01}

\begin{eqnarray}
P_{D}&=&e^{-\mu} \sum_{n=3}^{\infty}\ \frac{\mu^n}{n!}
4|c_{min}(n)|^2\\
&=&1-e^{-\mu}
\bigg(\sqrt{2}\sinh{\frac{\mu}{\sqrt{2}}}+2\cosh{\frac{\mu}{\sqrt{2}}}-1\bigg).\nonumber
\end{eqnarray}

\bibliographystyle{apsrev}

\begin{thebibliography}{99}
\bibitem{loss} The total loss in $dB$ of the quantum channel is given by
$-10\log_{10}\eta_{t}$.
\end{thebibliography}

\end{document}